\documentclass[]{piparticle-final}
\usepackage{graphicx}
\usepackage{amsmath,amssymb}

\usepackage{cite} 
\usepackage{epstopdf}

\begin{document}

\volume{5}               
\articlenumber{050007}   
\journalyear{2013}       
\editor{S. A. Grigera}   
\reviewers{E. M. Forgan, School of Physics \& Astronomy, \\ \mbox{}\hspace{40.5mm} University of Birmingham, U.K.}  
\received{25 January 2013}     
\accepted{10 October 2013}   
\runningauthor{P. Esquinazi}  
\doi{050007}         

\title{Invited review: Graphite and its hidden superconductivity}

\author{P. Esquinazi\cite{inst1}\thanks{E-mail: esquin@physik.uni-leipzig.de}
        }

\pipabstract{
We review experimental results, from transport to magnetization measurements, on different
graphite samples, from bulk oriented graphite, thin graphite films to transmission electron microscope lamellae,
that indicate the
existence of granular superconductivity at temperatures above 100~K. The   accumulated evidence
speaks for a localization of the superconducting phase(s) at certain interfaces  embedded in
semiconducting crystalline regions with Bernal stacking order.
}

\maketitle

\blfootnote{
\begin{theaffiliation}{99}
\institution{inst1} Division of Superconductivity and Magnetism,
Institute for Experimental Physics II, Fakult\"at f\"ur Physik und
Geowissenschaften, Universit\"at Leipzig, Linn\'estrasse 5,
D-04103 Leipzig, Germany.
\end{theaffiliation}
}

\section{Introduction}
Over the past decade, our  interpretation of the magnetic and
transport properties of ordered graphite bulk samples has
experienced a  change respect to the partially accepted
general description of their intrinsic properties. The
description of graphite one finds in the not-so-old
literature tells us that it is a kind of (semi)metal
with a finite Fermi energy and carrier (electron plus hole)
densities per graphene layer at low temperatures $n_0 \sim 10^{10}
\ldots 10^{12}~$cm$^{-2}$, see e.g., \cite{mcc64,kelly,gru08}.
However, real samples are not necessarily ideal, we
mean defect-free, and therefore those carrier densities are not
necessarily intrinsic of ideal graphite. The exhaustive
experience accumulated in gapless and narrow band semiconductors
\cite{tsi97} already indicates us how important defects and impurities
(not necessarily magnetic ones but, for example, hydrogen)  are in
determining some of the measured properties. Therefore, taking
experimental data of real samples as intrinsic, without knowing
their microstructure and/or defect concentration, was  indeed a
misleading assumption in the past. This assumption has
drastically influenced the description of the band structure of graphite
 we found nowadays
in several books and publications. For example, if  graphite has a
finite Fermi energy $E_F$ (whatever the majority carriers are),
 as assumed everywhere, up to seven free
parameters have to be introduced \cite{kelly,dil77,sch09} to  describe
the apparently ideal band structure of Bernal graphite with the
well-known ABAB stacking order of the graphene layers.

The impact of well defined two-dimensional interfaces inside
 graphite samples \cite{bar08,ina00}
 had not been realized until recent studies of the transport properties as a function
of thickness of the graphite sample provided a link to the
microstructure of the samples obtained by transmission electron
microscope (TEM) studies. We also have to add the sensitivity of
the graphite transport properties  to very small amount of defects
\cite{arn09}. Those results \cite{bar08,arn09} do not only indicate
us that at least a relevant part of the carrier densities measured
in graphite is not intrinsic but also that the  metallic-like
behavior of the electrical resistance does not reflect ideal,
defect-free graphite \cite{gar12}. An anomalous vanishing of  the
amplitude of the Shubnikov-de Haas (SdH) oscillations decreasing
the thickness of the graphite samples was published,
more than 10 years ago, \cite{oha01} without attracting the
necessary attention, although those results already suggested that
the SdH oscillations are probably not intrinsic of the graphite
structure. These results are supported by the absence of SdH
oscillations, i.e., no evidence for the existence of a Fermi
surface,  found recently in bulk oriented samples of high grade
and high purity but without internal interfaces \cite{cam12}.
 All these results indicate that
the internal microstructure of the graphite samples play an important role, a microstructure that
was neither characterized nor  considered in the discussion of the
measured properties of different graphite samples, from highly oriented pyrolytic graphite (HOPG) to Kish or natural graphite,
even in nowadays literature \cite{sch09,orl08JP,gon12}.

What does this have to do with superconductivity? If we  start
searching for superconductivity in graphite by measuring the
behavior of the electrical resistance ($R$) with temperature ($T$)
and magnetic field ($H$), for example, it should be clear that the
knowledge of the  intrinsic, normal state dependence is needed.
Otherwise, we may misleadingly interpret an anomalous behavior due
to, for example,  the influence of non-percolative, granular
superconducting regions embedded in a (normal state) graphite
matrix, as intrinsic of the  material, clearly missing an
interesting aspect of the sample. A reader with expertise in
superconductivity might not be convinced that such a mistake could
be ever made. However,  the ballistic transport characteristics of
the graphene layers in ideal graphite with their huge mobility and
mean free path \cite{gon07,gar08,dus11,esq12} provide a  high
conductivity path in parallel; such that it is not at all
straightforward  by simple experiments  to realize and prove the
existence of  superconductivity  at certain regions in some, not
all, graphite samples. One needs indeed to do  systematic
experiments decreasing the size of the graphite samples (but not
too much) to obtain clear evidence for the embedded or ``hidden"
superconductivity.

\begin{figure}[t]
\begin{center}
\includegraphics[width=0.45\textwidth]{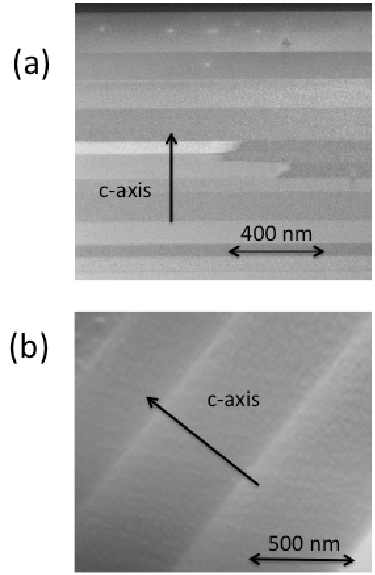}
\end{center}
\caption{Transmission electron microscope pictures of two
different kinds of interfaces and their distribution in HOPG
samples. The TEM pictures were taken from two different lamellae,
each about 300~nm thick and with the electron beam nearly parallel
to the graphene planes of the samples.  (a) The interfaces are
recognized at the borders of crystalline regions
 of different gray colors. Taken from \cite{bar08}. (b) Interfaces found in a HOPG sample used for magnetization
 measurements (see section~V)
 that reveals hysteretic behavior in field and temperature.
 Taken from \cite{schcar}.} \label{tem}
\end{figure}

A note on samples: The internal ordering or mosaicity of the graphite crystalline regions  inside
commercial HOPG samples is given usually by  the grade. For example, the
 highest ordered pyrolytic graphite samples have is a grade ``A",
 which means a rocking curve width $\Delta \sim 0.4^\circ \pm 0.2^\circ$
 (``B", $\Delta \sim 0.8^\circ$, etc.). Interestingly, and due to the contribution of
 two dimensional highly conducting
 internal interfaces between crystalline regions
 \cite{bar08,gar12}, the highest grade, i.e., smaller rocking
 curve width,  does not always mean that
 the used sample provides the intrinsic transport properties of ideal graphite.
 The characterization of the internal structure of usual HOPG samples, as well as the
 thickness dependence of $R(T)$  to understand the transport and the magnetic
 properties of graphite, indicate that these two dimensional interfaces are
 of importance. The existence of rhombohedral inclusions \cite{lin12,lui11} (stacking order ABCABC instead
 of ABAB of the usual Bernal graphite structure) in HOPG as well as in Kish graphite
 samples can also have a relationship with the hidden superconductivity
 in graphite, following the
 theoretical work in Ref. \cite{kop13}.
  According to literature (see e.g., Fig.~2-2 in Ref.~\cite{ina00}), the density of interfaces
 parallel to the graphene layers in Kish graphite, in
regions of several microns length, is
 notable. Therefore, quantifying the perfection of any graphite sample
 through the resistivity ratio between 300~K and 4.2~K  \cite{ina00}
 is not necessarily the best criterion to be used if we are interested on the intrinsic properties of the graphene
 layers in graphite, because of  the high conductivity of the interfaces in parallel to the
 graphene layers of the sample \cite{gar12}. Two examples of the interfaces we are referring to can be seen in Fig. \ref{tem}. On the other hand,
 commercial HOPG bulk samples are
 of high purity with
 average total impurity  concentrations below 20~ppm. Especially the existence of
 magnetic impurities are
 of importance if the Defect-Induced Magnetism (DIM) is the main research issue. Their
 concentration remains  below a few ppm for high grade HOPG samples \cite{jems08}.

The graphite flakes discussed in this work were obtained by exfoliation of HOPG samples
of different batches, by
careful mechanical
press and rubbing the initial material on a previously
cleaned substrate. As substrate, we used p-doped Si
with a 150 nm SiN layer on top. We selected the flakes using microscopic
and micro-Raman techniques to check their quality. More details on the preparation
can be taken from Ref.~\cite{bar08} and other publications cited below.

This review is organized as follows. In the next section we
discuss the experimental data for $R(T,H)$ from different graphite
samples published in the last 12 years and argue that the first
hints on unusual superconducting contribution can be already
found  in those measurements. In section ~III, 
 we
discuss the anomalous hysteresis in the magnetoresistance, a first
clear indication for embedded granular superconductivity.
Section~IV, 
deals with the Josephson behavior measured in
TEM lamellae whereas section~V 
deals with the granular
superconducting behavior found in the magnetization of
water-treated graphite powder as well as in bulk HOPG samples for
fields normal to the interfaces found inside those samples. In the
last section, section~VI
, before the conclusion, we discuss possible
origins for the superconducting signals on the basis of earlier
and recent experimental and theoretical work.

\section{The behavior of the resistance vs. temperature
at different applied magnetic fields}\label{RTs}

\begin{figure*}[t]
\hspace{+1cm}
\includegraphics[width=1.1\textwidth]{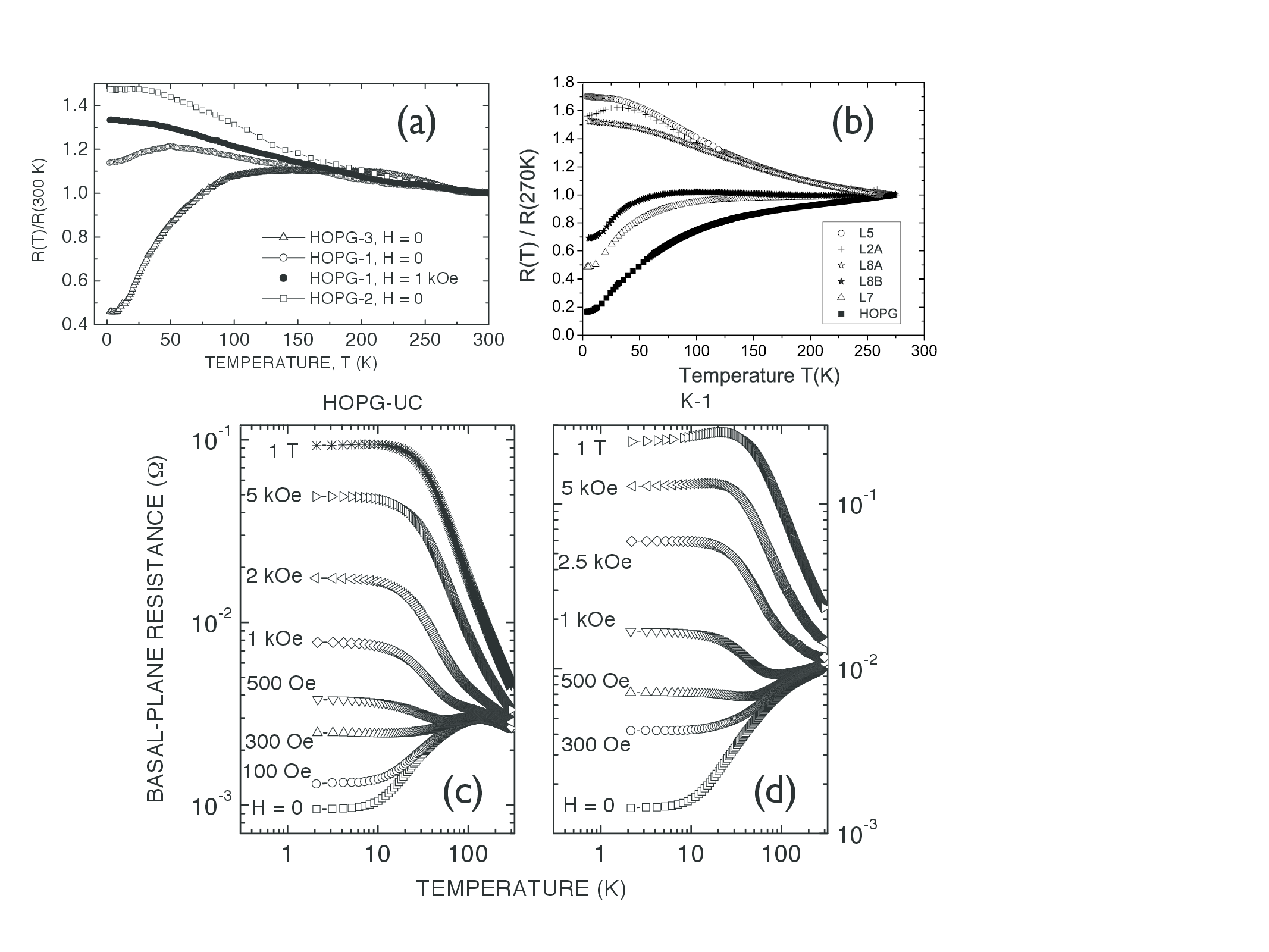}
\caption{(a) Normalized resistance vs. temperature for three
different HOPG bulk samples. The bottom, metallic-like curve
corresponds to the sample HOPG-3, the curves above correspond to
HOPG-1 ($H=0$), HOPG-1 ($H=1~$kOe), and HOPG-2. The grade
and resistivity values are HOPG-1 ($\Delta = 1.4^\circ$,
resistivity at 300~K $\rho(300$K$)= 45~\mu\Omega$cm), HOPG-2
($\Delta = 1.2^\circ$, $\rho(300$K$)= 135~\mu\Omega$cm) and HOPG-3
($\Delta = 0.5^\circ$, $\rho(300$K$)= 5~\mu\Omega$cm). Taken from
~\cite{kempa00}. (b) Similar to (a) but for HOPG samples from the
same batch but with different size, namely (thickness $\times$
length $\times$ width) L5: $12 \pm 3~$nm, $27~\mu$m, $14~\mu$m,
L2A: $20 \pm 5~$nm, $5~\mu$m, $10~\mu$m, L8A: $13 \pm 2~$nm,
$14~\mu$m, $10~\mu$m, L8B: $45 \pm 5~$nm, $3~\mu$m, $3~\mu$m, L7:
$75 \pm 5~$nm, $17~\mu$m, $17~\mu$m, HOPG: $17 \pm 2~\mu$m,
$4.4~$mm, $1.1~$mm, taken from ~\cite{bar08}. (c) and (d)
Resistance of bulk graphite samples vs. temperature at different
applied fields normal to the graphene layers. The sample in (c) is
a HOPG bulk sample from Union Carbide of grade A and the sample in
(d) is Kish graphite. Taken from ~\cite{yakovadv03}.} \label{RT}
\end{figure*}

In this section, we discuss the behavior of the  resistance
$R(T,H)$ of different HOPG samples including Kish graphite. The
data we present here  were taken from
\cite{bar08,kempa00,yakovadv03} and a quick search in literature
demonstrates that these data are reproducible and can be found in
different publications, see, e.g., \cite{kelly,tok04,heb05,kim05}.
Figure~\ref{RT}(a) shows the $R(T)$ for different bulk graphite
samples of different grades (rocking curve width) and for one
sample (HOPG-1) at  zero and under a magnetic field applied normal
to the main area, i.e., normal to the graphene planes of the
sample. This figure reveals a general behavior, namely that the
lower the resistivity $\rho$ of the HOPG sample, the more
metallic-like its temperature dependence. It is  appealing to
assume that these characteristics, low $\rho$, low $\Delta$ and
the metallic behavior are  clear signs for more ideal graphite.
Therefore, from the measured $R(T)$, we may conclude that sample
HOPG-3 is more ideal than sample HOPG-1 and the latter being more ideal than
sample HOPG-2, see Fig.~\ref{RT}(a).  This is indeed the usual
interpretation found in several reviews in the literature, see
e.g.,~\cite{del01,ina00}.

From a quick look at all the curves in Fig.~\ref{RT}, however,
one recognizes a striking similarity between them,
although we are comparing
different samples with different thickness and some of the curves were measured under a
magnetic field applied normal to the graphene layers of the samples; also normal to the
interfaces commonly found in some ordered samples \cite{ina00,bar08}.

Let us start discussing the metallic-like behavior of $R(T)$ of
sample HOPG-3 in Fig.~\ref{RT}(a). This sample behaves as the
HOPG-UC sample shown in (c) at zero magnetic field, having also a
maximum at $T \sim 150~$K. A ``better" metallic character shows
the HOPG sample in (b) or the Kish graphite sample in (d), without
any maximum in the shown temperature range. Is this metallic-like
behavior really intrinsic of ideal graphite?

The following experimental evidence  does not support such
interpretation:\\ First, for samples from the same batch, the
metallic character of $R(T)$ vanishes in the whole temperature
range when the sample thickness is below $\sim 50~$nm
\cite{oha01,bar08,kim05}, see Fig.~\ref{RT}(b).\\ Second, the
metallic-like behavior vanishes in the whole temperature range
after applying a magnetic field of the order of 1 to 2~kOe, an
interesting behavior known as the Metal-Insulator Transition (MIT)
\cite{kempa00,heb05,tok04}, see Figs.~\ref{RT}(a), (c) and (d).
Note that such magnetic field strength influences mainly the
metallic-like region,  see e.g., the change of sample HOPG-1 in
Fig.~\ref{RT}(a) at zero and at 1~kOe field, an interesting
behavior noted first in~\cite{yakov99} and interpreted as due to
superconducting instabilities.

At those field strengths, i.e., $H \sim 1~$kOe, the obtained $R(T)$
curves, for samples showing at zero field a metallic-like behavior,
resemble the semiconducting-like curves obtained for sample
HOPG-2 (Fig.~\ref{RT}(a)) or for samples with small thickness
(Fig.~\ref{RT}(b)). At fields higher than a few kOe, the rather
large magnetoresistance of graphite starts to play the main role
and the
 $R(T)$ curve increases in the whole temperature range.

 Finally, all these results added to  the existence of well defined
 interfaces in the metallic-like HOPG  samples as well as in Kish graphite,  with
distances in the $c-$axis direction usually larger than $\sim 30$~nm, indicate that the
metallic-like behavior is due to the contribution of these interfaces and it is
not intrinsic of the graphene layers of ideal graphite \cite{bar08,gar12}.
Therefore, explanations of the MIT based on
 ideal graphite band models with a large number
 of free parameters \cite{tok04,heb05} are certainly not the appropriate
 ones.

All the different $R(T)$ curves for different samples shown in
Fig.~\ref{RT} and at zero field can be very well understood
assuming the parallel contribution of semiconducting graphite
paths in parallel to the one from the highly conducting interfaces
\cite{gar12}. The saturation of the resistance at $T \rightarrow
0$~K is interpreted as due to the finite resistance of the sample
surfaces (the free one and the one on the substrate) short
circuiting the intrinsic behavior of the bulk graphene layers at
low enough temperatures.

The question is now whether parts of these interfaces hide
superconducting regions. It is certainly appealing to suggest that
the huge MIT at fields normal to the interfaces (and below $\sim
2~$kOe) is related to Josephson-coupled superconducting regions
embedded in some of the interfaces.  Note that the huge anisotropy
of the MIT (parallel fields to the interfaces main plain do not
affect the electrical transport) already implies that the regions
responsible for the MIT must be laying parallel to the graphene
layers \cite{kempa03}. Without the knowledge on the existence of
these interfaces, an  interpretation of the low field MIT based on
the influence of superconductivity has been discussed in detail in
the reviews \cite{yaknar,yakovadv03}. In those reviews, one can
recognize the remarkable similarity between the scaling approaches
used to characterize the magnetic-field-induced
superconductor-insulator quantum phase transition \cite{fis00} or
of the field-driven MIT in 2D electron (hole) systems \cite{abr01}
and the one obtained for the MIT observed in graphite. As we will
see in the next sections, the experimental evidence obtained in
the last recent years indicates that granular superconductivity
exists within some of those interfaces, indeed.

If superconducting patches exist embedded in parts of the interfaces or in other two dimensional regions of the
 bulk ordered samples, one
 expects to measure some signs of granular superconductivity, as for example
 nonlinear $I-V$ curves or hysteresis in the magnetoresistance.  However, this is not really observed in
 bulk large samples. There are at least two  reasons for the apparent
 absence of these expected phenomena.
 One is
 the distribution of the input current between the
  ballistic channels
 given by the graphene layers \cite{dus11,esq12}, the metallic, normal conducting parts of the
 interfaces  and the regions where the superconducting
 patches exist. In other words, the
 usual maximum currents used in transport experiments reported in bulk samples
 may have been
 small enough so that
 the current through the superconducting regions remained below the critical Josephson one. The other reason is
 the experimental voltage sensitivity to measure the possible irreversibility in the magnetoresistance due to the existence of
pinned vortices or fluxons.
We will see in the next section that part of these problems can be overcome decreasing
the sample size; in this way, one obtains the voltage signals from the regions of interest  with enough
sensitivity.

Apart from the large magnetic field sensitivity of the metallic-like resistance
measure in bulk graphite samples with interfaces, is there any further hint for the existence of
granular superconductivity in those
$R(T)$ curves? Yes, this hint is related to the thermally activated function ($\propto \exp(-E_a/k_BT )$ with $E_a$, a sample dependent effective thermal barrier $\sim 30~$K) one needs
in order to
fit the metallic-like contribution below $T \sim 200~$K \cite{gar12}. This function
is relevant in spite of only a factor five increase of the resistance between low and high temperatures, see Fig.~\ref{RT}.
Skeptical readers can convince themselves about its relevance  taking a similar example, as
 the exponential function used to fit the
increase, by a similar factor, of the ultrasonic attenuation with
temperature below $T_c$ in conventional superconductors. We note
that this exponential function has already been used to describe
the increasing resistance of bulk graphite samples with
temperature and it was speculated to be related to some
superconducting-like behavior in graphite \cite{yaknar}. It is
clear that this function is not the usual one, one expected for
metals or semimetals and that cannot be understood within the usual
electron-phonon interaction mechanisms, nor in two dimensions.  A
similar dependence has been observed in granular AlGe
\cite{sha83}, which shows for a particular Al concentration a
superconductor-semiconductor transition similar to that reported
in Ref. \cite{gar12} or, after an appropriate scaling in temperature,
to some of the curves shown in Fig.~\ref{RT}. The observed
thermally activated behavior might be understood on the basis of
the Langer-Ambegaokar-McCumber-Halperin (LAMH) model
\cite{lan67,mcc70} that applies to narrow superconducting channels
in which thermal fluctuations can cause phase slips. This
interpretation gets further support from the evidence we discuss
in the following sections.

\section{Hysteresis in the magnetoresistance}\label{hys}

In order to reveal by transport measurements the existence of
granular superconductivity in some regions of the graphite
samples, we need to increase the sensitivity of the measured
voltage to those regions. To achieve this, we decrease the size of
the sample enhancing in this way the probability to get some
measurable influence of this phenomenon in the voltage. The work
in Ref.~\cite{esq08} reported the first observations of an
anomalous irreversible behavior in the magnetoresistance (MR) in a
few tens of nm thick and several micrometer large multigraphene
samples. Hysteresis in the magnetoresistance is a key evidence on
the existence of either magnetic order (domains with their walls,
for example) or vortices/fluxons and therefore on the existence of
superconductivity. Because defects as well as hydrogen can trigger
magnetic order in graphite, a first attempt would be to relate the
measured hysteresis in the MR with the existence of magnetic order
and magnetic domains, for example. However, the data exhibited
anomalous hysteresis loops in the MR \cite{esq08}, similar to
those observed in granular superconductors with Josephson-coupled
grains \cite{ji93,kope01,fel03}. The anomalous hysteresis was
observed only for magnetic fields perpendicular to the planes,
whereas in the parallel to the planes direction, the MR remains
negligible. This fact already points out to a remarkable large
anisotropic response of the superconducting phase(s) in agreement
with the hypothesis that these superconducting regions might be
embedded in some of the interfaces found inside some bulk graphite
samples \cite{bar08,ina00}. The amplitude of the hysteresis in the
MR reported in Ref. \cite{esq08} vanishes at temperatures $T \sim
10~$K, clearly below the temperature at which  the resistance
shows a  maximum, as it is the case for samples HOPG-1 in
Fig.~\ref{RT}(a) or sample L2A in Fig.~\ref{RT}(b).

It is clear that thermal fluctuations can prevent the establishment
of a coherent superconducting state in parts of the sample
and therefore zero resistance state is not so simple to be achieved if the superconducting
distribution is a mixture of superconducting patches at the interfaces and these are embedded in
a multigraphene semiconducting matrix. Moreover, we should take also into account that the
voltage electrodes are usually connected at the top surface of the graphite samples
picking the voltage difference coming from a
non-negligible normal conducting path.

\begin{figure*}[th]
\begin{center}
\includegraphics[width=0.9\textwidth]{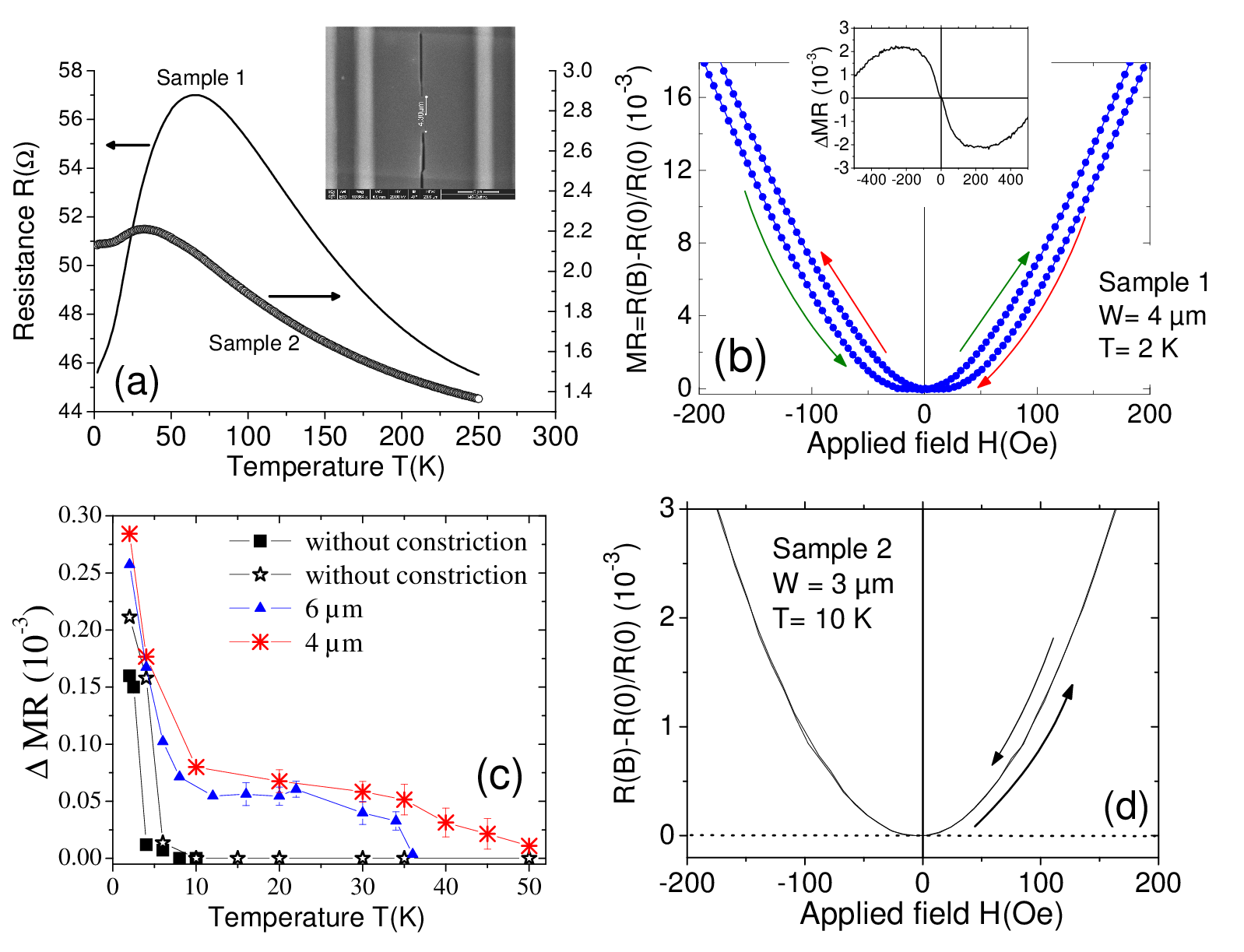}
\end{center}
\caption{(a) Resistance vs. temperature,
without constrictions and at zero applied field, for two graphite flakes of size
(distance between voltage electrodes $\times$ width $\times$ thickness)
for sample 1 (2): $13 \times 16 \times 0.015~ (2.6 \times 6 \times 0.040)$ $\mu$m$^3$. The observed temperature
dependence remains for all constrictions widths. The inset shows
a scanning electron microscope picture of sample 1 with a constriction
width of 4.3~$\mu$m between the two voltage electrodes. The scale bar is
5$~\mu$m.  (b) Magnetoresistance (MR) vs. applied magnetic field for sample 1 with
a $4~\mu$m constriction width  and at 2 K. The input current was $1~\mu$A.
Note the clear hysteresis in the MR when the field is swept from $|H_{\rm max}| = 1000~$Oe.
The inset shows the difference $\Delta$MR between the curve obtained starting
from $H_{\rm max} = +1000~$Oe
and the return curve measured from $H_{\rm min} = -1000~$Oe.
(c) The absolute difference
between the two MR curves
of the hysteresis loop obtained at a
fixed magnetic field of $16.6$~Oe for sample~1  without constriction $(\star)$ and for two different
constriction widths. The figure also shows the corresponding data for another graphite sample
without constrictions $(\blacksquare)$ from \cite{esq08}. (d) Magnetoresistance
measured from a starting
maximum field of 1.4~kOe at
10~K  for sample 2 with a
constriction width of $3~\mu$m.
Some of the figures and the data were taken from \cite{sru11}.} \label{hy}
\end{figure*}

One possibility to increase the sensitivity of the measured voltage
to the field hysteresis these regions produce
 is to
make  a constriction in the middle of the two voltage electrodes, see inset in Fig.~\ref{hy}(a).
In this case, we expect  a locally narrower distribution of
superconducting and normal regions at the constriction such that averaging effects should be
less  important.
Simultaneously, through the constriction the main part of the
voltage drop depends mostly on the region at the constriction, see Fig.~2(c) in Ref.~\cite{gar08}. Then, a higher sensitivity to the superconducting
paths  can be achieved in case they remain at or near the constriction. This idea has been successfully
realized in ~\cite{sru11}
and its main results will be reviewed in this section.

Let us take two slightly different samples, 1 and 2, with  $R(T)-$curves as shown
in Fig.~\ref{hy}(a). The aim of the experiment is  to study the hysteresis in the MR
 those samples might show below the temperature at which a maximum in the resistance is
measured,  in case that maximum is
 related to the Josephson coupling between superconducting regions. Figure~\ref{hy}(b)
shows one example of the anomalous hysteresis loop in the MR. The
going down curve (from high, positive to low, negative fields, red
arrow), for example, runs below the going  up curve (green arrow)
in the same quadrant as the field sweep was started, showing a
minimum at positive fields of the order of 20~Oe, see also similar
curves in Ref.~\cite{esq08}. To present the anomalous behavior
clearly, the inset in Fig.~\ref{hy}(b) shows the difference
between the two curves. This difference is  in clear contrast to
the usual hysteresis
 in superconductors as
 well as ferromagnets \cite{ji93,esq08},
 where the minimum (or maximum) in the MR is
 observed in the opposite field quadrant, and the increasing
 field resistance curve is usually below the decreasing field one.

Figure~\ref{hy}(c) shows the temperature dependence of the
difference in the MR between the decreasing and increasing field
curves at a fixed magnetic field for sample~1, without and with
two constrictions. The results show that the smaller the
constriction width, the higher the temperature at which the
anomalous hysteresis is observed, decreasing below the sensitivity
limit at $T > 50$~K for a constriction width of $4~\mu$m, whereas
the maximum in the $R(T)$ curve is at $\sim 70~$K, see
Fig.~\ref{hy}(a). The absence of any hysteresis in the MR for
sample~2 with a constriction width of $3~\mu$m and at $T = 10~$K
indicates that the hysteresis does not come from some artifact due
to the used focused ion beam method \cite{sru11,barnano10} or due
to an artifact in the measurement of the real field applied to the
sample. As expected from the $R(T)$ curve, see Fig.~\ref{hy}(a),
sample~2 shows the anomalous hysteresis in the MR at lower
temperatures than sample~1 \cite{sru11}.

Summarizing this section, the observation of the anomalous
hysteresis in the MR -- together with the MIT and the relatively
large MR at temperatures below the maximum in $R(T)$ -- provides
already  striking hints that granular superconductivity is at work
in some regions of these samples. The increase in the temperature
region where the hysteresis is observed, decreasing the
constriction width, demonstrates the problem of current averaging
and voltage sensitivity limits usual experiments with large
samples have.

\begin{figure*}[th]
\begin{center}
\hspace{-2cm}
\includegraphics[width=1.1\textwidth]{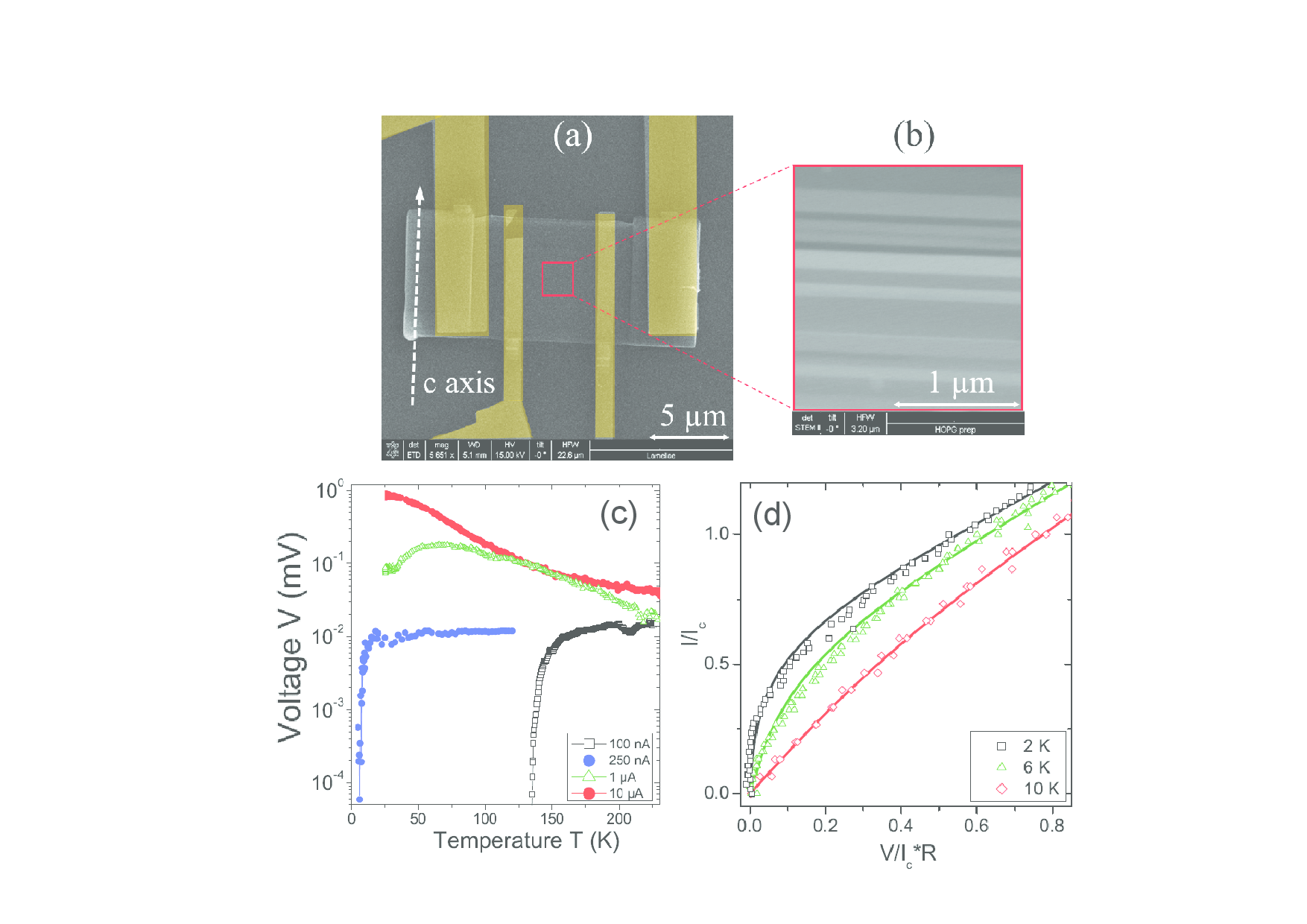}
\end{center}
\caption{(a) Scanning Electron Microscopy (SEM) image of a lamella
of 300~nm thickness on a Si/SiN substrate where the yellowish
colored areas are the electrodes. A four-point configuration has
been prepared with the outer electrodes used to apply current and
the inner ones to measure the voltage drop. The $c-$axis runs
parallel to the substrate surface and normal to the current
direction. (b) Transmission Electron Microscopy (TEM) image of a
HOPG lamella. The  different brightness corresponds to a different
orientation within the $a-b$ plane of the crystalline regions with
thickness $ > 30$~nm. (c) Voltage vs. temperature at different
input currents for a lamella of $\sim 800$~nm thickness and with Van der Pauw contact configuration. (d)
Current-Voltage characteristics at different temperatures for a
lamella of $\sim~300$~nm thickness  in reduced coordinates, where
$R$ is the normal state resistance, $I$ the input current, and
$I_c$ the critical Josephson current.   The continuous curves are
fitted to the model proposed in Ref. \cite{amb69} with $I_c(T)$ as the
only free parameter.  Figures taken from \cite{bal13}.} \label{la}
\end{figure*}

\section{Direct evidence for Josephson behavior
in the transport properties of graphite: Measurements in TEM lamellae}\label{lam}

If the embedded  interfaces (or some other quasi two dimensional regions) inside
 the measured graphite samples have superconducting properties, the best way to check them would be  contacting
 electrodes as near as possible to those interfaces or interface regions
 and study the behavior as a function of any useful parameter
 one can take to influence their response. It should be clear that one cannot simply open
 the graphite sample at the interface and  put voltage electrodes at the open surfaces of the
 interface, simply because
 it will not remain anymore.  A tentative  approach to put the contacts as near as
 possible to an interface has been done
 in Ref.~\cite{barjsnm10-s}. Indeed, the observed behavior at low temperatures and as a function of magnetic field
 appeared to be superconducting-like.

 A better and appealing evidence for the superconducting behavior
 embedded in some graphite ordered samples
 can be obtained by trying to locate the voltage electrodes directly at the inner edges of the interfaces. The work in Ref.~\cite{bal13} prepared TEM lamellae from bulk HOPG samples and using lithography
and focused ion beam techniques,  current and voltages electrodes
at different positions of the samples were prepared. In this way, one tries to  contact  several of those
 interface edges simultaneously, as shown in Fig.~\ref{la}(a).
 We note, however, that a thin surface layer of disordered graphite exists due to the Ga$^+$ ion irradiation
 used to cut the lamella from the bulk HOPG sample. This layer has a much larger resistance than the
 one of the graphene layers or of the interfaces \cite{barnano10}
  and  therefore the input current goes through the lowest resistance  path
  as well as the voltage electrodes pick up
 the response of the graphite sample with its interfaces. One can see this
 comparing first  the $R(T)$ curves obtained
 at large enough currents in the lamellae (Fig.~\ref{la}(c))
 with those of graphite samples with top electrodes (Fig.~\ref{RT}).
 The fact that a  zero resistance state (minimum voltage noise $\pm$ 5 nV upon sample) is obtained at low currents
with $I-V$ characteristic curves that resemble the
 one expects for Josephson coupled grains leaves little doubt about the origin of the obtained signals.
 In the TEM picture of Fig.~\ref{la}(b), one can see the
 graphite single crystalline regions (different gray colors) oriented differently between them
 about the common $c-$axis
  and having well defined two dimensional interfaces,
 as high resolution TEM studies revealed \cite{anajov}.

Figure \ref{la}(c) shows the voltage vs. temperature measured in a
TEM lamella of oriented graphite \cite{bal13} at different input
DC currents, from 100~nA to 10$~\mu$A. The clear sharp transition,
observed at $\sim 150$~K at the lowest current, shifts to lower
temperatures increasing the input DC current. For the largest
input currents, the temperature dependence of the resistance of
the contacted lamella shows a maximum or
follows the intrinsic semiconducting behavior of the graphene layers.
This behavior already suggests
the existence of high temperature granular
superconductivity at some parts of the sample. The study reported
in Ref. \cite{bal13} shows that the transition temperature depends on
the prepared sample. This indicates a sample dependent
distribution of the superconducting regions and/or some influence
of the preparation process or sample size on the superconductivity \cite{anajov}.
We also note that the observed sharp decrease in the measured
voltage does not necessarily indicate the critical temperature of
the superconducting regions but the temperature below which a
percolative granular system shows negligible resistance due to the
Josephson coupling  at the used input current.

Current-voltage characteristic curves at different temperatures
and in different lamellae obtained from different HOPG samples
have been studied in Ref. \cite{bal13}. An example of this $I-V$ curves
at three temperatures is shown in Fig.~\ref{la}(d) obtained for a
different lamella. The curves follow the expected dependence for a
Josephson junction \cite{amb69} with a temperature dependent
critical current, the only free parameter in the fit.

Further evidence that speaks  for a superconducting origin of the
$I-V$ curves is given by the expected detrimental effect of a
magnetic field on the superconducting state. This effect can be
due to an orbital depairing effect  or due to the alignment of the
electron spins at much higher fields, in case of singlet coupling.
The effect of a magnetic field applied normal and parallel to the
interfaces has been studied in detail for thick and thin lamellae
in Ref. \cite{bal13}. Upon sample size (thickness, i.e., width of the
graphene planes inside  the lamella) the observed effects are from
the usual vanishing of the zero resistance state or no effect at
all for thin lamellae. A magnetic field of a few kOe applied
normal to the interfaces is enough to destroy the Josephson
coupling at low temperatures, an effect compatible with the MIT
observed in several graphite samples, see section II. Whereas a
field applied parallel to them does not influence the $I-V$ curves
at all, a fact that speaks for the two dimensionality of the
superconducting regions.

Nevertheless, the influence of a magnetic field in HOPG samples
with interfaces is not as ``simple" as in conventional
superconductors. For high fields applied normal to the interfaces,
the $I-V$ curves show a recovery to the zero resistance state. The
observed reentrance appears to be related to the magnetic-field
driven reentrance observed at low temperatures  in the longitudinal
resistance at high enough magnetic fields \cite{yakovprl03}. This
interesting behavior as well as the insensitivity  of the $I-V$
curves to magnetic fields in very thin lamellae \cite{bal13}
deserve further studies.

We would like to note here that the
possible effects of a magnetic field on the superconducting state
of quasi two-dimensional superconductors, or in case the coupling
does not correspond to a singlet state, are not that clear as in
conventional superconductors. For example, results in two
different two-dimensional superconductors, including one produced
at the interfaces between non superconducting regions
\cite{gard11}, show that superconductivity can even be enhanced by
a parallel magnetic field. In case the pairing is
$p-$type \cite{gon01}, the influence of a magnetic field is expected
to be qualitatively different from the conventional, singlet
coupling behavior \cite{sch80,kni98}  with even an enhancement of
the superconducting state at intermediate fields. In case the
London penetration depth is much larger than the size of the
superconducting regions at the interfaces of our lamellae or if
the superconducting coherence length is of the order or larger
than the thickness of the lamella, the influence of a
magnetic field should be less detrimental.

Through these studies, and taking into account that in samples
without these  interfaces no signature
of superconducting or metallic-like behavior has been observed (see also 
section V) it is appealing to suggest that
superconductivity is somewhere hidden at some of those interfaces
or interface regions. It should be also clear that not all those
interfaces have superconducting regions  with similar critical
parameters. Those interfaces are formed during the preparation of
the HOPG samples based on treatments at very high temperatures ($T
> 3400^\circ$C) and high pressures ($P \sim $10~kg/cm$^3$) and in
a non-systematic way \cite{ina00}. Actually, they are not at all an
aim of the production but actually the opposite, they should be
avoided in order to enhance the crystal perfection of the bulk
HOPG material. It is even possible that, upon the procedure used
to control the structure and texture of the graphite sample, the
near surface region, for example, can have a different degree of
graphitization as inside the bulk HOPG sample \cite{ina00}. This
means that one may obtain different results from different parts
of the same HOPG sample. Therefore, disconcerting situations and
an apparent lack of reproducibility are preprogrammed in case the
research studies are done without taking care of the internal
microstructure of the studied samples.

\section{Magnetization measurements }\label{M}

In this section, we present and discuss  magnetization
measurements done in bulk HOPG samples with and without embedded
interfaces and in water treated graphite powders. One of the main
problems in interpreting  magnetization data for fields applied
parallel to the $c-$axis of the graphite structure, i.e., normal to
the graphene layers and interfaces, is the need of subtraction of
a large diamagnetic background. Due to the small amplitude of the superconducting-like signals
in the studied samples, the subtraction of this linear in field background is not
so simple, because it is not known  with enough certainty to obtain
the true field hysteresis after its subtraction. That means that
we always have a
certain arbitrariness in the shape of the obtained field hysteresis, a
situation that  will improve with the increase of the amount of
 material responsible for those superconducting-like signals.
The small SQUID signals of interest imply that one should  take
additional efforts to minimize or rule out
 possible SQUID artifacts \cite{pas91,ney08,saw11}. Therefore, systematic studies
of samples of different or equal geometry and magnetic background, with and without
interfaces, are necessary.

Taking into account: the overall shape of the hysteresis, the
slope of the virgin curve at low fields where the subtraction does
not affect too much, and the overall experience with
ferromagnetic graphite \cite{jems08,barapl11}, one can rely to a
certain extent on the obtained  hysteresis shape. Certainly, not
only the field hysteresis but also other evidence one gets from
magnetization measurements  as, e.g., the remanence at zero field
as a function of the maximal field applied (see for example
measurements for YBa$_2$Cu$_3$O$_7$ in Ref. \cite{mce90}) and the
hysteresis in temperature dependent measurements helps to convince
oneself about the existence of some kind of granular
superconductivity. The hysteresis between the field cooled (FC)
and zero-field cooled (ZFC) curves can help to discern between a
superconducting or ferromagnetic-like behavior. The most obvious
evidence that speaks against a simple ferromagnetic order of the
hysteresis observed as a function of temperature and field is the two dimensionality of the obtained hysteretic
signals \cite{schcar}
, i.e., the superconducting-like
signals are mainly measured for fields normal to the interfaces.
This fact is not compatible with any kind of magnetic order
including shape or magneto crystalline anisotropy, whatever large
they might be. We note that the ferromagnetic response of graphite
due to DIM is mostly measured for fields parallel to the graphene
layers, parallel to the main area of the samples \cite{jems08}.

\subsection{Bulk graphite samples}

\begin{figure}[t]
\begin{center}
\includegraphics[width=0.45\textwidth]{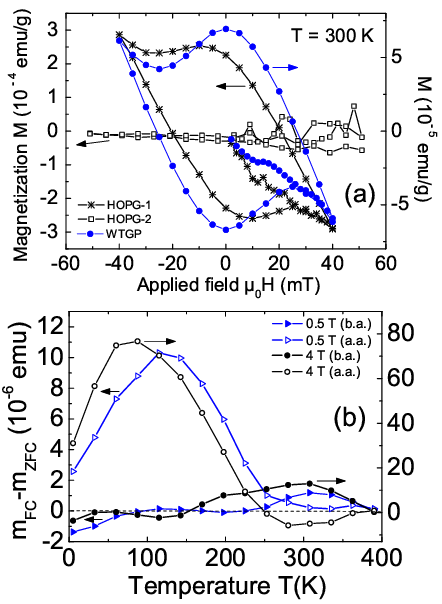}
\end{center}
\caption{(a) Magnetization of two HOPG bulk samples (HOPG-1 and
HOPG-2) after subtraction of a diamagnetic background  and of
water treated graphite powder (WTGP, right $y-$axis) at 300~K. The
HOPG-2 sample shows no hysteresis in contrast to the other two
samples. (b) Temperature dependence of the difference between FC
and ZFC magnetic moments of the HOPG-1 sample  before (b.a.) and
after (a.a.) warming the sample up to $\simeq 600~$K, at two
constant applied fields, 0.5~T (left $y-$axis) and 4~T (right
$y$-axis). The field was applied always normal to the interfaces
or graphene planes of the samples. Data taken from
\cite{schcar}.} \label{bulk}
\end{figure}

Figure~\ref{bulk}(a) shows the field hysteresis, after subtraction
of the corresponding diamagnetic linear background, at 300~K of
two bulk HOPG samples, HOPG-1 and HOPG-2 and a water treated
graphite powder (WTGP) (right $y-$axis). A TEM characterization of
the internal microstructure of the HOPG-1 sample shows clear
evidence for well defined interfaces running parallel to the
graphene layers, in contrast to the HOPG-2 sample \cite{schcar},
see Fig.~\ref{tem}(b). These results clearly indicate that the
origin of the hysteresis is related to the existence of the
interfaces in the HOPG-1 sample. The absence of the hysteresis in
the HOPG-2 sample, which has a  similar diamagnetic background and
overall geometry as the HOPG-1 sample, also indicates that the
hysteresis is not due to an obvious SQUID artifact or an artifact
in the background subtraction. The field hysteresis is similar to
that of WTGP. The narrowing of the hysteresis observed at high
fields is expected for granular superconductors
\cite{sen91,bor91,and01,sch12}. From the hysteresis, as well as
measuring the remanent magnetic moment as a function of the
applied field \cite{schcar}, one obtains the characteristic
Josephson critical fields $h_{c1}^J(T)$ and $h_{c2}^J(T)$  with
values similar to the WTGP \cite{sch12} and a similar ratio
$h_{c2}^J(T)/h_{c1}^J(T) \sim 3$ \cite{schcar}.

Figure~\ref{bulk}(b) shows the magnetic moment hysteresis in
temperature (FC minus the ZFC curve) for the HOPG-1 sample as
received (b.a.) and after sweeping the temperature up to 500~K
(a.a.) \cite{schcar}, at  two applied fields.  We would like to
stress the following features: The hysteresis for the
as-received sample starts from the turning point (390~K) and it is
positive. The hysteresis in temperature at both applied fields
are qualitatively similar, showing a crossing to negative values
at low temperatures. Larger ZFC values (smaller in absolute value)
than FC ones in the magnetic moment are usually not observed,
neither in superconductors nor in ferromagnets and it does appear
to be a SQUID artifact \cite{schcar}. This negative hysteresis in
temperature would suggest that the superconducting properties can
be enhanced to some extent under a magnetic field, an effect that
might be related to the reentrance we have shortly mentioned in
section IV. A slight annealing of the HOPG-1 sample of less
than one hour at $\sim 500$~K changes drastically the observed
hysteresis for both fields (open symbols in Fig.~\ref{bulk}(b)).
The hysteresis appears to be shifted to lower temperatures but
with negative values at high temperatures and high fields. We note
that annealing at similar temperatures for several hours produced
a decrease in the overall hysteresis observed in WTGP (see
supporting information of Ref.~\cite{sch12}). At the state of this
research, it is unclear whether pinning properties of vortices
and/or of fluxons or the existence of different superconducting
phases play a main role in the  hysteresis that is observed for
fields applied normal to the interfaces.

\subsection{Water treated graphite powder}

\begin{figure}[t]
\begin{center}
\includegraphics[width=0.45\textwidth]{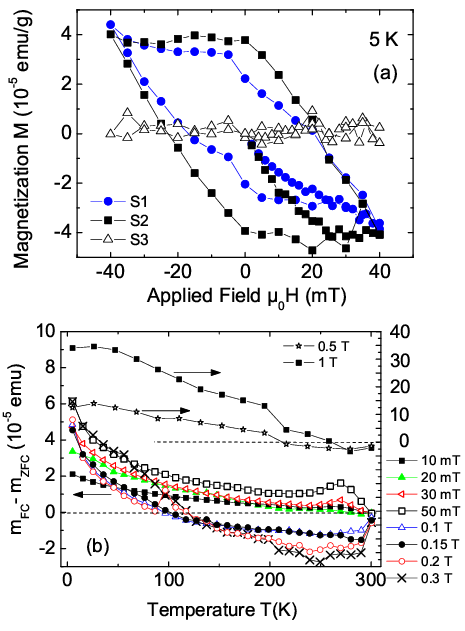}
\end{center}
\caption{(a) Field hysteresis at 5~K for a maximum applied field
of 40~mT for the water treated graphite powder  (S1), the same
powder but after pressing it in a pellet with a pressure of $18
\pm 5$~MPa (S2) and after pressing it again with a pressure of $60
\pm 20$~MPa (S3). The corresponding diamagnetic linear backgrounds
were subtracted from the measured data. (b) Difference between the
FC and ZFC curve at different applied fields for a water treated
graphite powder. Data taken from Ref.~\cite{sch12}.} \label{pow}
\end{figure}

The work of Ref.~\cite{sch12} reports on the magnetic response of
WTGPs. The main message of that work is
that the WTGP shows a hysteretic behavior in field and temperature
compatible with granular superconductivity. As an example, we show
in Fig.~\ref{pow}(a) the field hysteresis at 5~K of a WTGP (S1,
lose powder without applying significant pressure) and the same
WTGP but after pressing it into a pellet with two different
pressures (S2,S3). After the diamagnetic background subtraction,
the field hysteresis is similar to that obtained for bulk HOPG
sample with interfaces, see Fig.~\ref{bulk}(a) for similar data
but at 300~K. The fact that the hysteresis vanishes after applying
pressure to the powder rules out simple SQUID artifacts
(the diamagnet background does not diminish after making a pellet
from the graphite powder, but the contrary) and also it rules
out that the hysteresis is due to a ferromagnetic response due to impurities.

Figure~\ref{pow}(b) shows the difference in the magnetic moment
between the ZFC and FC curves, as in Fig.~\ref{bulk}(b). The
behavior of this difference as a function of the applied field
appears to be compatible with the one expected for granular
superconductors \cite{sch12}. Note the following features: The
hysteresis increases at all $T$ for fields $\mu_0H \lesssim
50~$mT, showing a maximum near the turning point of 300~K, similar
to the HOPG-1 sample in the as-received state, see
Fig.~\ref{bulk}(b). At fields 0.1~T $ \lesssim \mu_0 H \lesssim
0.2~$T the difference decreases at all $T$ and remains rather
field independent. At higher fields, however, it increases showing
a shift of the crossing point (from negative to positive values)
to higher $T$ . This behavior is at odds to the one expected for
ferromagnets, even for ferromagnetic nanoparticles \cite{pro99} as
well as for superconductors with a pinning force that decreases
with applied field in the shown field range. From the results in
\cite{sch12},  and using basic concepts of vortex pinning, we would
then conclude that if an upper critical field exists, then it
should be clearly larger than 7~T in the temperature range of the
figure.

In spite of some interesting differences between the behavior
obtained for bulk HOPG and WTGP, the similarities already suggest
that the water treatment helps to produce a certain amount of
interfaces between graphite grains, being the
origin for the whole hysteresis. Thermal annealing as well as
pressing the WTGP are detrimental indicating that defects and/or
hydrogen or oxygen at the interfaces could play an important role in the
observed phenomena.

\section{Discussion}\label{dc}

Superconductivity in carbon-based systems is a rather old, well
recognized fact. This phenomenon was probably first observed in
the potassium intercalated graphite $C_8K$ \cite{han65} back in
1965. Since then, a considerable amount of studies reported this
phenomenon in carbon-based systems, reaching critical temperatures
$T_c \sim 10$~K in intercalated graphite \cite{wel05,eme05} and
above 30~K - though not percolative - in some HOPG samples
\cite{yakovjltp00} as well as in doped graphite and amorphous
carbon systems \cite{silvaprl,kop04,fel09,kopejltp07}. Traces of
superconductivity at $T_c = 65~$K have been recently reported in
amorphous carbon powder that contained a small amount of sulfur
\cite{fel12}. Superconductivity was found also in carbon nanotubes
with $T_c = 0.55~$K~\cite{koc01} and 12~K \cite{tak06} or possibly
even higher critical temperatures \cite{tan01,zhao01}.
Superconductivity with $T_c \sim 4~$K in boron-doped diamond
\cite{eki04} and in diamond films with  $T_c \sim 7~$K
\cite{tak04} belong also to the recently published list of
carbon-based superconductors. We should note, however, that
superconductivity at room temperature in a disordered graphite
powder has been already reported in 1974 \cite{ant74}, see also
\cite{ant75}, a work that did not attract the necessary attention
in the community.

Whether quasi two dimensional interfaces play a role in the above
mentioned carbon-based superconductors, one can probably rule out
only for the intercalated graphite and doped diamond compounds,
where the three dimensional superconductivity is characterized by
a relatively low critical temperature. We may speculate that the
traces of superconductivity found in doped amorphous carbon,
disordered or ordered graphite powders may be related to some
interfaces between well ordered graphite regions. The experience
of the high temperature superconducting oxides already suggests
that two dimensionality is advantageous to achieve higher critical
temperatures.

Apart from the usual transport and magnetization measurements used
to characterize the superconducting state, there are scanning
tunneling spectroscopy (STS) results obtained on certain
disordered regions of a HOPG  surface at $T = 4.2~$K that revealed
an apparent energy gap $\sim 100~$meV \cite{agr92}. Although the
overall curves resemble a superconducting-like density of states,
the authors suggested that the gap originates from charging
effects. See further STS results and the discussion in
\cite{kopejltp07}.

Theoretical works that deal with superconductivity in graphite as
well as in graphene have been published in recent years. For
example, $p$-type superconductivity has been predicted to occur in
inhomogeneous regions of the graphite structure \cite{gon01} or
$d-$wave high-$T_c$ superconductivity \cite{nan12} based also on
resonance valence bonds \cite{doni07}, or at the graphite surface
region with rhombohedral stacking due to a topologically protected flat band \cite{kop11}.

For the graphite structure, the experimental evidence obtained in the last years suggests
that high temperature superconductivity exists at certain interfaces or
interface regions within the usual Bernal structure
although the structure of the superconducting
regions remains unknown.
 One can
further speculate that  due to the high carrier concentration that
can be localized at those interfaces, they should be
predestined to play a role in triggering superconductivity.
Following a BCS approach in two dimensions (with anisotropy), for
example, a critical temperature $T_c \sim 60~$K has been estimated
if the density of conduction electrons per graphene plane
increases to $n \sim 10^{14}~$cm$^{-2}$, a density that might be
induced by defects and/or hydrogen ad-atoms \cite{garbcs09} at the
interfaces, or by Li deposition \cite{pro12}. Further predictions
for superconductivity in graphene support the premise that $ n >
10^{13}~$cm$^{-2}$ in order to reach $T_c
> 1~$K \cite{uch07,kop08}. On
the other hand, the possibility to have high temperature
superconductivity at the surface of or in the rhombohedral
graphite phase \cite{kop11,kop13} -- a phase that sometimes  is
found in graphite samples \cite{lui11,lin12} -- stimulates further
careful studies of these hidden interfaces. In the last years,
superconductivity has been found at the interfaces between oxide
insulators \cite{rey07} as well as between metallic and insulating
copper oxides  with  $T_c \gtrsim 50~$K\cite{goz08}. Also,
interfaces in different Bi bicrystals
 show superconductivity up to 21~K, although Bi
bulk is not a superconductor \cite{mun06,mun08}.

Finally, we think that some of the interfaces are  also the origin
for the metallic-like behavior of graphite samples as well as for
the quantum Hall effect (QHE) found in some HOPG samples
\cite{yakovprl03,yakadv07}. Because the existence, density as well
as the intrinsic properties of these interfaces depend on sample,
we can now understand why the reproducibility  of the QHE in bulk
HOPG samples is rather poor.

\section{Conclusion}\label{c}
In this review, we have discussed the following experimental evidence:\\
Firstly, the temperature and magnetic field dependence of the electrical resistance of bulk and thin films of
graphite samples
and its relation with the existence of two dimensional interfaces. \\
Secondly,  the Josephson behavior of the current-voltage
curves with an apparent zero resistance state at  high temperatures in especially made TEM lamellae.\\
Thirdly, the anomalous hysteresis in the magnetoresistance observed in graphite thin samples as well as its enhancement
restricting the current path within the sample. \\
Finally, the overall magnetization of bulk graphite samples,
with and without interfaces, as well as water treated graphite
powders.

All this experimental evidence as a whole indicates the existence
of superconductivity located at certain interfaces inside graphite
samples. Although we cannot rule out other interpretations for
some of the observations discussed in this work, the whole
evidence suggests that superconductivity should be the origin for
all the phenomena discussed here. Clearly, the situation is still
highly unsatisfactory because several open questions remain,
namely, the characteristics of the superconducting phase(s), from
the structure to the main superconducting parameters, as ``simple"
as the critical temperature and critical fields, the coherence and
penetration lengths, etc. It is clear that further studies are
necessary in the future but the overall work done until now shows
us the way to go.

\begin{acknowledgements}
The author acknowledges the support provided by the Deutsche
Forschungsgemeinschaft under contract DFG ES 86/16-1 and the
ESF-Nano under the Graduate School of Natural Sciences
``BuildMona". The results presented in this review were part of
the Ph.D. thesis of Heiko Kempa (section II),  Srujana Dusari (section
III) and Ana Ballestar (section IV) as well as  the master thesis of
Thomas Scheike (section V) done in the Division of Superconductivity
and Magnetism of the Institute for Experimental Physics II of the
University of Leipzig. The author thanks  Dipl. Kris. Annette
Setzer, Dr. Jos\'e Barzola-Quiquia and Dr. Winfried B\"ohlmann for
their experimental assistance and support. The permanent support
as well as the discussions  with Nicol\'as Garc\'ia are gratefully
acknowledged. Special thanks go to Yakov Kopelevich with whom we
started in the year 1999 and in a rather naive way the research of
a new and unexpected world behind graphite.
\end{acknowledgements}

\section*{Note added in proof} Since the submission of this
manuscript, some new works related to the subject of this review
were published. Tight-binding simulations done in Ref. \cite{mun13}
support the work done in Ref. \cite{kop13} and found that surface
superconductivity is robust for ABC stacked multilayer graphene,
even at very low pairing potentials.  Through the observation of
persistent currents in a graphite filled ring-shaped container
immersed in alkanes, the author in Ref. \cite{kaw13} claimed possible
room temperature superconductivity. For completeness, we include
them in the reference list.

\bibliographystyle{pip}


\end{document}